\documentclass[a4paper]{mem}
\usepackage{natbib}
\usepackage{graphicx}
\usepackage[a4paper]{hyperref}
\idline{73}{1}
\begin{document}
   \title{X--ray and optical coverage of \\
   3EG J0616--3310 and 3EG J1249--8330}
   \author{N. La Palombara \inst{1}, G.F. Bignami \inst{2,3}, P. Caraveo \inst{1}, \\
   A. De Luca \inst{1}, S. Mereghetti \inst{1}, R. Mignani \inst{4}, E. Hatziminaoglou \inst{5}}
   \offprints{N. La Palombara}
\mail{via E. Bassini 15, 20133 Milano}

   \institute{IASF/CNR  - Sezione di Milano `G.Occhialini', Via E. Bassini 15, I-20133 Milano (I) \email{nicola@mi.iasf.cnr.it} \\
              \and CESR, 9 Avenue du colonel Roche, F-31028 Toulouse (F) \\
              \and Universit\`a di Pavia, Dipartimento di Fisica Teorica e Nucleare, Via Ugo Bassi 6, I-27100 Pavia (I) \\
              \and European Southern Observatory, Karl Schwarzschild Strasse 2, D-85740 Garching (D) \\
              \and Instituto de Astrofisica de Canarias, Via Lactea, E-38200 La Laguna-Tenerife (E) \\
             }

   \abstract{
The limited angular resolution of $\gamma$--ray telescopes prevents the straight identification of
the majority of the sources detected so far.  This is particularly true for the low latitude,
probably galactic ones, only 10 \% of which has been identified.  Most of the counterparts are {\it
Isolated Neutron Stars}, both radio--loud and radio--quiet, characterised by an extremely high value
of $f_{X}/f_{opt}$.  The best way to search for INSs in the error boxes of unidentified EGRET sources
is to perform the X--ray coverage of the $\gamma$--ray field, followed by the optical
characterization of each X--ray source.  We applied this procedure to two EGRET sources, which could
belong to a local galactic population:  3EG J0616--3310 and 3EG J1249--8330.  Here we report on the
analysis of about 300 X--ray objects, as well as on their optical study.
   }
   \authorrunning{N. La Palombara et al.}
   \titlerunning{3EG J0616-3310 and 3EG J1249-8330}
   \maketitle
%
%________________________________________________________________

\section{Introduction}

The third EGRET catalogue (\citet{Hartman}) lists 271 high--energy
$\gamma$--ray sources: the high--latitude ones ($|b|\ge 10^{\circ}$), presumably
extragalactic, are 183, while the remaining 88 are at low latitudes ($|b|\le
10^{\circ}$) and should belong to our Galaxy.

Source identification has been hampered by several problems:  the EGRET poor angular resolution
($\simeq1^\circ$); the source confusion; the variety of potential emitters; the low count
statistics.  Up to now, only about 100 sources have been identified:  67 are blazars, 27 are
candidate blazars, while only 7 sources have been associated to $\gamma$--ray pulsars.  Therefore
about 170 EGRET sources are still unidentified.  In particular, this is true for more than 90\% of
the low--latitude, presumably galactic ones (\citet{Caraveo 2001}).

At low/medium latitude Isolated Neutron Stars (INSs) are the most promising candidates as
$\gamma$--ray source counterparts.  Up to now, INSs have provided the only confirmed
identifications, while various attempts to associate the low--latitude unidentified EGRET
sources with different classes of galactic objects have not yielded conclusive results
(\citet{Romero}; \citet{Geherels}).

Unfortunately, the identification of an INS is very difficult:  the limited statistics prevents the
direct detection of a $\gamma$--ray time signature and may render difficult the search for an
already known radio signal.  The situation is even worse for radio--quiet INSs, where the absence of
a reference radio pulsation makes it even more difficult to detect a significant time signature.

In order to search for both {\em radio--loud} and {\em radio--quiet} INSs we can take advantage of
one of their characteristics:  the extremely high value ($>$100) of the X--ray--to--optical ratio
($f_{X}/f_{opt}$).  Thus, INSs can be found through a 2--step procedure:
\begin{enumerate}
\item each $\gamma$--ray error--box is covered by dedicated X--ray observations, which will
yield an harvest of new X--ray sources
\item the error--box of each X--ray source is scrutinized in the optical waveband, in
order to find its counterpart or absence thereof and to single out potential neutron stars using the
$f_{X}/f_{opt}$ values
\end{enumerate}

\begin{figure}
\centering
\resizebox{\hsize}{!}{\includegraphics[angle=+90,clip=true]{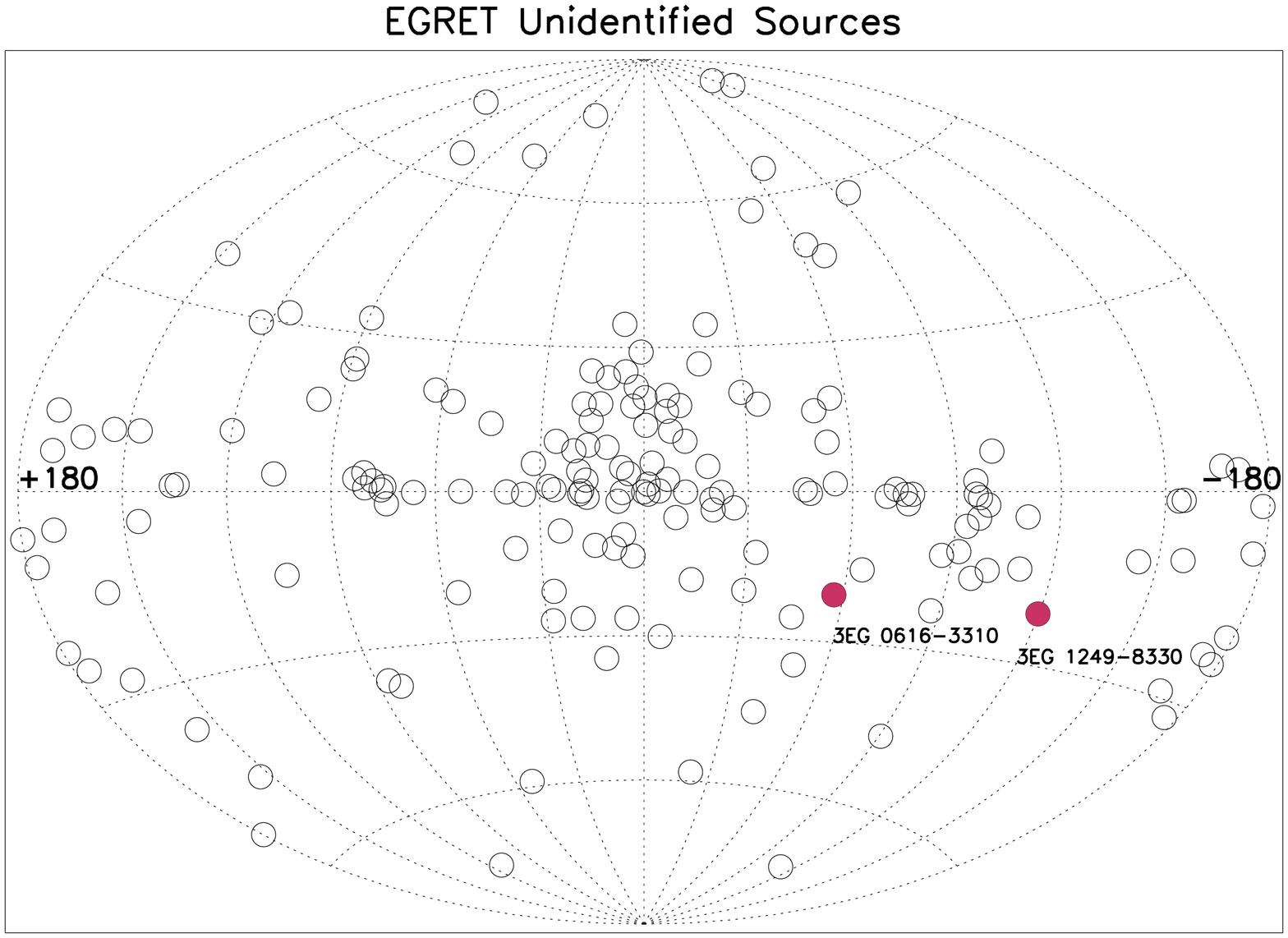}}
\caption{Sky  distribution  of  the  unidentified  {\em EGRET}  sources:  the two red circles show the celestial  position of 3EG0616--3310
and 3EG1249--8330 (courtesy of S.Vercellone)}\label{galacticposition}
\end{figure}

Such an investigation strategy was devised for the identification of {\em Geminga}
(\citet{Bignami}) and is being used for the study of other EGRET sources (\citet{Mirabal};
\citet{Caraveo 2001}).  In this work we applied it to two middle--latitude EGRET sources:  3EG
J0616--3310 ($l=240^\circ$ 20$'$ 45.4$''$, $b=-21^\circ$ 14$'$ 31.7$''$) and 3EG J1249--8330
($l=302^\circ$ 51$'$ 35.2$''$, $b=-20^\circ$ 37$'$ 44.8$''$).  They have been selected owing to
their relatively good positional accuracy, spectral shape, galactic location and lack of
candidate extra--galactic counterparts.  In Fig.~\ref{galacticposition} we outline their
celestial position compared to the other unidentified {\em EGRET} sources.

\section{X--ray observations and results}

The error--boxes of 3EG J0616--3310 and 3EG J1249--8330 are circles of $\simeq0.5^\circ$
radius. Thus it was
possible to cover each error--box with four $\simeq$ 10 ks observations of the
{\em XMM--EPIC} focal plane camera. In each observation the {\em pn} camera (\citet{Struder}) was operated in {\em Extended Full Frame} mode, while the {\em MOS1} and
{\em MOS2} cameras (\citet{Turner}) were operated in standard {\em Full
Frame} mode; in all cases the {\em thin} filter was used.

After the standard processing pipeline, we performed the source detection in 7 different energy
ranges:  namely we considered two coarse soft/hard bands (0.5--2 and 2--10 keV) and a finer
energy division (0.3--0.5, 0.5--1, 1--2, 2--4.5, 4.5--10 keV); we selected only the sources
with a {\it detection lihelihood} $L>8.5$ in at least one of our energy ranges.  We found a total of
146 sources in the 3EG J0616--3310 error box and 148 sources in the 3EG J1249--8330 one.

In Table~\ref{detections} we report the number, as well as the percentage, of sources detected in
each energy range (since a source can be detected in more than one energy band, the percentage
values do not add up to 100).

\begin{table}[h]
\caption{X--ray sources detected in each energy range.}\label{detections}
\begin{center}
\footnotesize{
\begin{tabular}{|c|c|c|} \hline
Source 		& J0616-3310	& J1249-8330	\\ \hline
Range (keV)	&	N(\%)	&	N(\%)	\\ \hline
0.5--2		& 119 (81.5)	& 119 (80.4)	\\
2--10		& 41 (28.1)	& 42 (28.4)	\\ \hline
0.3--0.5		& 28 (19.2)	& 14 (9.5)		\\
0.5--1		& 73 (50)		& 43 (29.1)	\\
1--2		& 81 (55.5)	& 77 (52)		\\
2--4.5		& 47 (32.2)	& 36 (24.3)	\\
4.5--10		& 4 (2.7)		& 8 (5.4)		\\ \hline
Total		& 146		& 148		\\ \hline
\end{tabular}}
\end{center}
\end{table}

From Table~\ref{detections} we can see that almost all of the sources are detected between 0.5 and 2
keV and more than 50 \% of them are also detected in the sub--range 1--2 keV; on the other hand,
only few sources are detected at very high or very low energies.  It is also interesting to outline
that, below 1 keV, the percentage of detected sources is lower for 3EG J1249--8330 field than for
3EG J0616--3310, whereas they are in full agreement at higher energies:  very probably this
difference is due to the column density of the interstellar gas, which is higher in the first field.

We used the count rates (CR) in the seven energy ranges to calculate the source {\em Hardness
Ratios}.  We found that our source population is characterised by rather soft spectra:  in all
the count distributions the peak is between 0.5 and 2 keV and, in the majority of the cases,
the CR is higher in the 1--2 keV range than in the 0.5--1 keV one.

\section{Optical analysis}

In order to perform a first search for the optical counterparts of our X--ray sources, we
cross--correlated them with two optical/infrared catalogues:  GSC2.2 (http://www--gsss.stsci.edu) and
2MASS (http://pegasus.phast.umass.edu).  In Tab.~\ref{statistics} we report the number of sources
with and without a candidate optical counterpart as well as the total number of candidate
counterparts.  In few cases we found more than one candidate optical counterpart; on the other hand,
half of the X--ray sources do not have a candidate counterpart.  This result comes as no surprise,
since the limiting {\it photographic} magnitude of the two catalogues is $F\simeq$ 20, while we
estimated that the visual brightness of the faintest counterparts should be $\simeq$ 22 for stars
and $\simeq$ 25 for other classes of objects.  For the X--ray sources with no counterpart we set $F$
= 20:  this magnitude has to be considered as the lower limit for any undetected counterpart of our
X--ray sources.

\begin{table}[h]
\caption{Results of the cross--correlation 
of the X--ray sources with the optical catalogues}\label{statistics}
\begin{center}
\begin{tabular}{|c|c|c|} \hline
EGRET Field	& J0616-3310	& J1249-8330	\\ \hline
Sources		&	146	&	148	\\ \hline
With no		&		&		\\
counterpart	&	71	&	74	\\ \hline
With		&		&		\\
counterpart	&	75	&	74	\\ \hline
Candidate		&		&		\\
Counterparts	&	97	&	95	\\ \hline
\end{tabular}
\end{center}
\end{table}

\begin{figure}
\centering
\resizebox{\hsize}{!}{\includegraphics[angle=-90,clip=true]{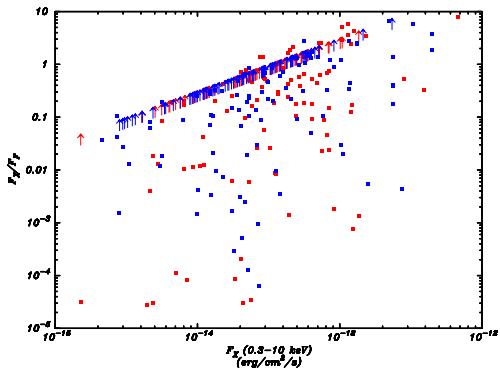}}
\caption{X--ray--to--optical   flux   ratio  for  the   3EG J0616--3310   (red)   and
3EG J1249--8330 (blue) fields as a function of the X--ray flux;  X--ray sources with no
optical counterpart are indicated with arrows.}\label{fluxratio1}
\end{figure}

From the available magnitude values (or from their limit) we derived the optical flux (or its
upper limit) of each source counterpart; then we calculated the X--ray--to--optical flux ratio
$f_{x}^{0.3-10}/f_{F}$.  For each source, we derived the X--ray flux from the net count rate,
taking into account the applicable count/energy conversion factor for a power-law spectrum with
photon index $\alpha=1.7$.  In Fig.~\ref{fluxratio1} we plot the ratio values, as a function of the
X--ray fluxes, for all the detected X--ray sources, both with and without a candidate
counterpart.

In the case of the 149 sources with at least one candidate counterpart, the plot shows that there is
no correlation between the two quantities:  at all the X--ray fluxes there is a large spread in the
X--ray/optical ratios.  Moreover, for both the EGRET fields the maximum value of the flux ratio is
less than 10:  this means that the sources could be AGN or BL Lac but the ratios are certainly not
compatible with compact counterparts.

Also in the case of the 145 sources with no candidate counterpart the estimated flux ratios are
rather low:  but here we must emphasize that these values have to be considered as {\em lower
limits}. The real magnitudes are certainly fainter than the value we used (F=20). For instance, a counterpart with F=25
would imply an $f_{x}^{0.3-10}/f_{F}$ ratio 100 times higher than our lower limit, thus becoming a
serious INS candidate.  Therefore it makes sense to search amongst suct still unidentified X--ray
objects the most promising counterparts of the two EGRET sources.

In order to perform a reliable identification of these X--ray sources, it is necessary to reach at
least V$\simeq$25 in the optical follow-up of the X-ray observation, with a dedicated multicolour
optical survey.  To this aim, we have used the {\em Wide Field Imager} of the ESO 2.2 m
telescope as the optical complement of the XMM data. In fact, the field--of--view of the WFI is
comparable to the EPIC one.  32 hours of observation have already been performed at this facility
and the data analysis is now in progress.

\section{Conclusions and perspectives}

Thanks to the {\em XMM-EPIC} observations, we have detected about 150 X--ray sources in both the
error--boxes of 3EG J0616--3310 and 3EG J1249--8330.  We have found that $\simeq$ 50\% of the
sources have no candidate optical counterpart, up to a limiting magnitude of $F \simeq$ 20.  Since
they are often high $f_{x}/f_{opt}$ objects, this sub--sample seems very promising in order to find
the counterpart of the two unidentified EGRET sources.

Using the newly acquired WFI data, we will proceed towards the classification of the optical objects
with an automatic alghorytm (\citet{Hatziminaoglou}) and to complete the optical identification of our X-ray sources.

\bibliographystyle{aa}

\end{document}